\documentclass{article}
\usepackage[dvips]{graphicx}

\def\bc{\begin{center}}                 \def\ec{\end{center}}
\def\beq{\begin{equation}}              \def\eeq{\end{equation}}
\def\bear{\begin{eqnarray}}             \def\eear{\end{eqnarray}}
\def\lb{\label}                         \def\ci{\cite}
\def\la{\langle}    \def\ra{\rangle}
         
\def\Dlt{\Delta}    \def\vphi{\varphi}
\def\rar{\rightarrow}         \def\lrar{\longrightarrow}
\def\stackrela#1#2{\mathrel{\mathop{#2}\limits_{#1}}}

\def\strel{\stackrela}

\begin{document}

\title{\hfill{\normalsize quant-ph/0307137}\\[3mm]
\bf On the position uncertainty measure on the circle}
\author{D.A. Trifonov\\
Institute for nuclear research,\\ 72 Tzarigradsko chaussee, 1784 Sofia,
Bulgaria} \maketitle   

\begin{abstract}
New position uncertainty (delocalization) measures for a particle on the
circle are proposed and illustrated on several examples, where the
previous measures (based on $2\pi$-periodic position
operators) appear to be unsatisfactory. The new measures are suitably
constructed  using the standard multiplication angle operator variances.
They are shown to depend solely on the state of the particle and to obey
uncertainty relations of the Schr\"odinger--Robertson type.\\[-2mm]

\end{abstract}


\section*{{\large 1. Introduction}}

Recently there is a renewed interest to the old problem of the
uncertainties and uncertainty relations for a particle on the circle
\ci{KR03,T03,KR02,FA01,GO98,KRP96}.
Due to the controversial commutation relation between the angle and
angular momentum operators most attention have been paid to position
operators that are invariant under translation $\vphi \lrar \vphi + a$,
$a\in I\!\!R$.  The relation $[\hat{l}_z,\hat{\vphi}] = -i$  stems from the
Dirac correspondence rule
\beq\lb{1}
\{f,g\} \lrar i[\hat{f},\hat{g}]
\eeq
between Poisson bracket $\{f,g\}$ of two classical quantities $f$ and $g$
and the commutator of the corresponding operators $\hat{f}$ and $\hat{g}$
for quantum observables. This rule is formally satisfied with
\beq\lb{2}
\hat{\vphi} = \vphi\quad {\mbox{ \small and}}\quad \hat{l}_z=-id/d\vphi.
\eeq
However on the eigenstates
\beq\lb{3}
\psi_m(\vphi) = \exp(im\vphi)/\sqrt{2\pi},\qquad m=0,\pm1,\dots,
\eeq
the above commutation relation breaks down, together
with the associated standard Heisenberg-Robertson uncertainty relation
$(\Dlt l_z)^2(\Dlt\vphi)^2\geq 1/4$.  Therefore  authors try to adopt
another position operator \ci{FA01,GO98,N67}, or even another definition
of the uncertainty  on the circle \ci{KR02}.

In this letter we provide an approach to the issue with minimal (in our
opinion) deviation from the standard commutation relation and standard
measure of uncertainty. The main idea has been sketched in \ci{T03}. Here
we develop it in greater detail, providing some proofs and further
examples. After a brief review of the properties of main previous position
uncertainty measures in section 2, two different new measures are
constructed and discussed in section 3. The new measures are constructed
using suitably the standard expressions of the first and second moments of
the angle variable, calculated by integration over $2\pi$ intervals. They
are of the form of positive state functionals, the values of which depend
solely on the state considered.

The terms position uncertainty measure and state delocalization measure
are used here as synonyms. The uncertainty measure of states is also
called measure of spread of corresponding wave functions (more precisely
of the corresponding probability distributions $p(\vphi)=|\psi(\vphi)|^2$).
It is worth noting that all uncertainty measures are maps of the infinite
dimensional state space into the positive part of the real line. It is
impossible in such a way to distinguish between all states.  Therefore
different measures should be considered not only as competitive, but as
complementary as well.

\voffset=-10mm

\section*{\large 2. A brief review of previous delocalization measures}

For a particle on the real line the standard measure of the position
uncertainty is given by the second moment $(\Dlt x)^2:= \la (x -
\la x\ra)^2\ra$  of the position operator $\hat{x}=x$, or
equivalently by the standard deviation $\Dlt x$. Mathematically both $\Dlt
x$ and $\la x\ra$ are one-to-one functionals on state space.  The quantity
$(\Dlt x)^2$ is also called variance, or dispersion,  of $x$ and is also
denoted as $Dx$ or $M^{(2)}x$. The variance of $x$ is regarded as a
measure of spread, or delocalization, of the state wave function
$\psi(x)$.  More precisely this is a measure of spread of the probability
distribution $p(x):=|\psi(x)|^2$.  Here the means $\la  x\ra$ and $\la
x^2\ra$ are calculated by integration with respect to $x$:\, $ \la x\ra =
\int x|\psi(x)|^2 dx $.\,

However in the case of angle operator $\hat{\vphi} = \vphi$ it was not
clear how to calculate and interpret the analogous quantity $\Dlt\vphi$,
since the operator $\hat{\vphi} = \vphi$ is not invariant under
translation $\vphi \rightarrow \vphi+2\pi$ (not $2\pi$-periodic), while
the wave functions $\psi(\vphi)$ are $2\pi$-periodic by definition.  This
trouble seems to be the main reason many authors to look for
$2\pi$-invariant position operators in order to  construct relevant
uncertainty measures on the circle.

The first such operators used probably were $\sin\vphi$ and $\cos\vphi$
\cite{N67}. The variances of these operators satisfy correct
inequalities \cite{N67}
\beq\lb{4}
(\Dlt l_z)^2(\Dlt\sin\vphi)^2 \geq |\la\cos\vphi\ra|^2/4,\qquad
(\Dlt l_z)^2(\Dlt\sin\vphi)^2 \geq |\la\cos\vphi\ra|^2/4.
\eeq
However one can see that the variances $(\Dlt\sin\vphi)^2$ and $(\Dlt
\cos\vphi)^2$ may take values greater than the corresponding one for the
uniform distribution $p_0(\vphi) = 1/2\pi = |\psi_m(\vphi)|^2$: in
$\psi_m(\vphi)$ one has $(\Dlt\sin\vphi)^2 = (\Dlt\cos\vphi)^2 = 1/2$): In
$\psi_c(\vphi) = (1/\sqrt{\pi})\cos\vphi$ one has $(\Dlt\cos\vphi)^2=3/4$,
$(\Dlt\sin\vphi)^2=1/4$, and in $\psi_s(\vphi) = (1/\sqrt{\pi})\sin\vphi$
these are interchanged --  $(\Dlt\cos\vphi)^2=1/4$,
$(\Dlt\sin\vphi)^2=3/4$.  The two states $\psi_c(\vphi)$ and
$\psi_s(\vphi)$ coincide under the shift $\vphi \rar \vphi\pm \pi/2$,
therefore it is reasonable to have coinciding (or close) measures of
spread for them, which should be less than those in the eigenstates
$\psi_m(\vphi)$. These deficiencies are partially removed by the
"uncertainty measure" \ci{N67} $(\tilde{\Dlt}\vphi)^2 =  (\Dlt\cos\vphi)^2
+ (\Dlt\sin\vphi)^2$, which can be written also in the forms
\beq\lb{5}
(\tilde{\Dlt}\vphi)^2 =  1 - \la\cos\vphi\ra^2 - \la\sin\vphi\ra^2
= 1 - |\la U(\vphi)\ra|^2, \qquad U(\vphi)=e^{i\vphi}.
\eeq
The quantity $\tilde{\Dlt}\vphi$ has been considered also in \ci{GO98}
and \ci{FA01}. In \ci{FA01} it was noted that $\tilde{\Dlt}\vphi$ has the
meaning of radial distance of the centroid of the ring distribution
$p(\vphi)$ from the circle line (and $\la\cos\vphi\ra^2
+\la\sin\vphi\ra^2$ is the squared centroid's distance from the center of
the circle -- see figure 1 in \ci{FA01}).
From (\ref{5}) and (\ref{4}) it follows that \ci{N67}
\beq\lb{6}
(\Dlt l_z)^2({\tilde\Dlt}\vphi)^2 \geq \frac 14(\la\cos\vphi\ra^2 +
\la\sin\vphi\ra^2).
\eeq
This uncertainty relation is approximately minimized in the canonical
coherent states (CS) $|\alpha,\beta\ra$ of the two dimensional oscillator with
large value of Re$^2\alpha+{\rm Re}^2\beta$ \ci{N67}.

However if one consider the quantity $(\tilde{\Dlt}\vphi)^2$, eq.
(\ref{5}), as a delocalization measure on the circle one encounters some
unsatisfactory results. For example, it produces the same maximal
delocalization (i.e. $\tilde{\Dlt}\vphi = 1$) for the eigenstates
$\psi_m(\vphi)$ of $\hat{l}_z$ and for all states $\psi(\vphi)$
with the property $|\psi(\vphi)|=|\psi(\vphi+\pi)|$. The centroid for
those $\pi$-periodic distributions
$|\psi(\vphi)|^2$ is in the center of the ring. On figure 1 graphics of
three $\pi$-periodic distributions are shown: uniform one $p_0(\vphi) =
1/2\pi = |\psi_m(\vphi)|^2$,
$p_s(\vphi)=|\psi_s(\vphi)|^2=\sin^2\vphi\,/\pi$ and $p_{s2}(\vphi) =
|\sin(2\vphi)|^2/\pi$. It is clear that
the localization of that distributions is quite different, and it is
desirable to have an uncertainty measure that distinguishes between them.

A rather nonstandard expressions for position and angular momentum
uncertainties for a particle on the circle were introduced and discussed
in \ci{KR02}:
\beq\lb{7}
\Dlt^2(\hat{l}_z) =
\frac 14 \ln(\la e^{-2\hat{l}_z}\ra\la e^{2\hat{l}_z}\ra),\quad
\Dlt^2(\hat{\vphi}) = -\frac 14 \ln|\la U(\vphi)^2\ra|^2.
\eeq
For a large sets of states these quantities obey the inequality
$\Dlt^2(\hat{l}_z) + \Dlt^2(\hat{\vphi}) \geq 1 $, the equality being
reached in the eigenstates $|\xi\ra$ of the operator
$Z=\exp(-\hat{l}_z+1/2)U(\vphi)$. The family of $|\xi\ra$ is overcomplete
and the states $|\xi\ra$ are called CS on the circle
\ci{DG93,KRP96,GO98,KR02}.

The functional $\Dlt^2(\hat{\vphi})$ was proposed as a position
uncertainty on the circle.  However this uncertainty measure was found
\ci{T03} to be not quite consistent with state localization: on CS $|\xi\ra$
it equals $1/2$, while on the visually worse localized states
$|\xi\ra-|\!-\!\xi\ra$ (Schr\"odinger cat states on the circle) it can
take rather less value of $0.33$ (see \ci{T03} and figure 2 therein).
On the above noted states $\psi_s(\vphi)$, $\psi_{s2}(\vphi)$ and
$\psi_m(\vphi)$ it takes values $0.346$, $\infty$, $\infty$. Thus it makes
distinction between  $\psi_s(\vphi)$ and $\psi_{s2}(\vphi)$ and
$\psi_m(\vphi)$, but identifies $\psi_{s2}(\vphi)$ with the uniform state
$\psi_m(\vphi)$ (see figure 1). Another unsatisfactory property of
$\Dlt^2(\hat{\vphi})$ is, that it takes the smaller value of $0.143$ on
the two-peak state $\psi_{s4}(\vphi) = (0.2+\sin^2\vphi)^2/N$, while on
the CS $|z\!=\!1\ra$ (one-peak state) it assumes the much larger value of
$0.5$ (see figure 2).

The authors of \ci{KR03,KR02} do not consider the above noted properties as a
deficiency of $\Dlt^2(\hat{\vphi})$ and in support of such opinion provide
\ci{KR03} the example of two step functions $\psi(x)$ and $\phi(x)$ on the
real line: $\psi(x)$ is different from $0$ in the interval $(-L/2,L/2)$
only (where it takes the value $1/\sqrt{L}$); $\phi(x)$ is different from
$0$ in the two smaller intervals $(-L/2,-L/4)$, $(L/4,L/2)$ (where it
takes the value $\sqrt{2/L}$) (see figure 1 in \ci{KR03}).  The standard
second moment $(\Dlt x)^2$  for $\psi(x)$ is lesser than that for
$\phi(x)$. However the authors of \ci{KR03} write "the state
$|\psi\ra$ is much worse localized on the interval $|x|<L/2$, than the
state $|\phi\ra$. In fact, we know that in the state $|\phi\rangle$ the
particle is not in the region $|x|<L/4$". My remark is that the step
function $\phi(x)$ can not be regarded as a one particle state on the
interval $|x|<L/2$ exactly due to the fact that $\phi(x) = 0$ in the
region $|x|<L/4$. Due to this fact particle never can jump from the left
region $(-L/2,-L/4)$ to the right one $(L/4,L/2)$, and vice versa.
Therefore this example can not be interpreted against the reliability
of $\Dlt x$ as an uncertainty measure on the real line.

\section*{\large 3. Generalized uncertainty measures based on the variance}

The state space of a particle on the circle consists of $2\pi$-periodic
square-integrable functions $\psi(\vphi)$. (In fact periodicity is up to a
phase factor). In view of this periodicity the scalar product of two
states $\psi_1(\vphi)$ and $\psi_2(\vphi)$ can be calculated by
integration with respect to $\vphi$ within any interval of length $2\pi$.
Since $\vphi\psi(\vphi)$ is no more periodic in $\vphi$ the standard
second moment $D\vphi \equiv (\Dlt\vphi)^2$ of $\vphi$ would naturally
depend on the interval of integration (here specified by the reference
point $\vphi_0$),
\beq\lb{8}
D\vphi =
\int_{\vphi_0-\pi}^{\vphi_0+\pi}(\vphi-\la\vphi\ra_{\vphi_0})^2|\psi(\vphi)|^2
d\vphi = D\vphi (\vphi_0),
\eeq
$$\la\vphi\ra_{\vphi_0} =
\int_{\vphi_0-\pi}^{\vphi_0+\pi}\vphi|\psi(\vphi)|^2 =
M\vphi(\vphi_0).\eqno{(8a)} $$
This $\vphi_0$-dependence of the standard moments of $\vphi$ is the
main reason authors to abandon $D\vphi$ and to look for other expressions
to simulate quantum position uncertainties on the circle or, equivalently,
the spread of the related periodic probability
distributions $p(\vphi)$. It turns out however, that the variance
(\ref{8}) could still be useful in construction of relevant uncertainty
measures.

First of all we note, that if one defines the $\vphi_0$-dependent covariance
$\Dlt l_z\vphi(\vphi_0)$ of $\hat\vphi$ and $\hat{l}_z$ as the real part of
the matrix element\,
$G_{l_z\vphi}:=
\la(\hat{l}_z-\la\hat{l}_z\ra)\psi|(\hat\vphi-\la\vphi\ra)\psi\ra$,\,
$\Dlt l_z\vphi(\vphi_0)={\rm Re}\,G_{l_z\vphi}(\vphi_0)$,\,
(where the means are taken by integration as in (\ref{8})), one obtains the
inequality (see also \ci{T03} and \ci{TChD})
\beq\lb{9}
D\vphi(\vphi_0)D\,l_z - (\Dlt l_z\vphi(\vphi_0))^2 \geq ({\rm
Im}\,G_{l_z\vphi}(\vphi_0))^2,
\eeq
which is a generalization of the Schr\"odinger (or
Schr\"odinger--Robertson) uncertainty relation \ci{SR30}. For a particle on
the real line the latter relation read $Dx\,Dp - ({\rm Cov}(x,p))^2 \geq
1/4$, where ${\rm Cov}(x,p)\equiv
\Dlt\,xp$ is the covariance of $\hat x$ and $\hat p$. The problem remains
however to define on the circle uncertainty (or delocalization, or
spread) measure $\Dlt^2_{|\psi\ra}\vphi$ of the state $|\psi\ra$ (or of
the distribution $p(\vphi)$) that {\it depends solely on
the state} $|\psi\ra$, and not on the limit of integration in (\ref{8}).
It turned out that this problem can be resolved by a suitable use of
$D\vphi(\vphi_0)$ due to the $2\pi$-periodic property of the
functional (\ref{8}),
\beq\lb{10}
D\vphi(\vphi_0+2\pi) = D\vphi(\vphi_0).
\eeq
The property (\ref{10}) can be easily proved, using the state periodicity
$|\psi(\vphi+2\pi)|=|\psi(\vphi)|$ and the definition of
$D\vphi(\vphi_0)$.  In fact one can show that all moments
$M^{(n)}\vphi(\vphi_0)=\la(\vphi-\la\vphi\ra)^n\ra$, $n=1,\ldots$, of $\vphi$
are $2\pi$-periodic in $\vphi_0$. In view of this periodic property the
$\vphi_0$- independent uncertainty measure can be defined in two different
ways: \\
(a) as an arithmetic mean of $D\vphi(\vphi_0)$ with respect to
$\vphi_0\in I_{2\pi}$, and \\
(b) as an extremal value \footnote{We
consider the minimal value only, since the maximal one may be greater than
that of the uniform distribution.} of $D\vphi(\vphi_0)$ in
$I_{2\pi}$, where $I_{2\pi}$ is any interval of length $2\pi$,
\begin{eqnarray}
(a)\quad \,_a\Dlt^2\vphi \,=\,
\frac{1}{2\pi}\int_{I_{2\pi}}D\vphi(\vphi_0) d\vphi_0,\lb{11}\\[2mm]
(b)\quad \,_b\Dlt^2\vphi \,=\,\, \strel{\vphi_0\in
I_{2\pi}}{\mbox{minimum}}D\vphi(\vphi_0).\lb{12}\quad
\end{eqnarray}
We introduce also the arithmetic mean squared covariance (by integration in
any $2\pi$ interval $I_{2\pi}$)
\beq\lb{13}
\,_a(\Dlt l_z\vphi)^2 = \frac{1}{2\pi}\int_{I_{2\pi}}
(\Dlt l_z\vphi(\vphi_0))^2\, d\vphi_0.
\eeq
Then taking into account eqs. (\ref{9}), (\ref{11})--(\ref{13}) and the
fact that minimum of $D\vphi(\vphi_0)$ is achieved at
some $\vphi_0=\vphi_{min}$, we arrive at two Schr\"odinger type
uncertainty relations ($\Dlt^2l_z = D\,l_z = (\Dlt l_z)^2$)
\beq\lb{14}
\,_i\Dlt^2\vphi\, \Dlt^2\,l_z - \,_i(\Dlt l_z\vphi)^2 \geq
\,_i({\rm Im}G_{l_z\vphi})^2,
\eeq
where $i = a,b$ and $\,_a({\rm Im}G_{l_z\vphi})^2$ is the arithmetic mean
of $({\rm Im}\,G_{l_z\vphi}(\vphi_0))^2$.  Thus both measures
$\,_a\Dlt\vphi$ and $\,_b\Dlt\vphi$ are supported by inequalities of the
type of Schr\"odinger uncertainty relation. It follows from this analogy
that the quantities $\,_i\Dlt^2\vphi$,\, $\Dlt^2\,l_z$, and $\,_i(\Dlt
l_z\vphi)^2$ could be regarded as (generalized) second moments of $\hat
\vphi$ and $\hat{l}_z$.

The examinations show that in a variety of examples the quantities
$\,_a\Dlt^2\vphi$ and $\,_b\Dlt^2\vphi$ behave as relevant position
uncertainty measures on the circle. Both measures distinguish between all
states presented on figure 1 and figure 2, their value for the uniform
distribution being greater than that for the other distributions. On the
states in figure 1 and figure 2 we have a satisfactory
arrangement of the spread measures, consistent with the visualized
localization. The values of $\,_b\Dlt^2\vphi$, for example, read
\beq\lb{15}
\,_b\Dlt^2\vphi|_{p_0(\vphi)}=\frac{\pi^2}{3}\, >
\,\,_b\Dlt^2\vphi|_{p_{s2}(\vphi)} =
3.16\, > \,\,_b\Dlt^2\vphi|_{p_s(\vphi)} = 2.79,\quad  ({\rm figure\,\,1}),
\eeq
\beq\lb{16}
\frac{\pi^2}{3}\, >\, \,_b\Dlt^2\vphi|_{p_{s4}(\vphi)} = 2.61 \,> \,
 \,_b\Dlt^2\vphi|_{p_{cs}(\vphi)} = 0.5,\quad  ({\rm figure\,\,2}).
\eeq
Compare the results (\ref{15}) and (\ref{16}) with the corresponding
values of measures (\ref{5}) and (\ref{7}). For example compare (\ref{16})
with $\displaystyle \Dlt^2(\hat{\vphi})|_{p_{s4}} = 0.346\, < \,
\Dlt^2(\hat{\vphi})|_{p_{cs}} = 0.5.$

There is a third invariantly defined state characteristic point on the
circle (the first two are the points, where
$D\vphi(\vphi_0)$ attains its extrema). This third
point is the {\it center of the packet} $p(\vphi)$, denoted
here as $\vphi_c$.  For a large set of distributions the center of the
packet $\vphi_c$ can be defined and determined as the angle of
the centroid of $p(\vphi)$. The cartesian coordinates of the centroid are
$x=\la\cos\vphi\ra$ and $y=\la\sin\vphi\ra$. We define the third
measure of spread of $p(\vphi)$ as $\,_c\Dlt^2\vphi$ \ci{T03},
\beq\lb{17}
\,_c\Dlt^2\vphi =  D\vphi (\vphi_0\!=\!\vphi_c),
\eeq
where $D\vphi(\vphi_0)$ is the second moment (\ref{8}).

The choice of $\vphi_0=\vphi_c$ in the limits of integration in (\ref{8}) was
wrongly interpreted in \ci{KR03} as introduction of a {\em definition\/}
of average values depending on the particular state. To reveal this
misinterpretation suffice it to recall that $\vphi_c$ is a characteristic
point of the distribution $p(\vphi)$, therefore of the state
$\psi(\vphi)$: the value of $\vphi_c$, and thereby the value of
$\,_c\Dlt^2\vphi$ and $\la\vphi\ra_{\vphi_c}$ are determined solely by the
state $|\psi\ra$. Thus $\,_c\Dlt^2\vphi$, first proposed in \ci{T03}, is a
correct positive functional of the state and may be examined as an
uncertainty measure.

A problem with the definition (\ref{17}) appears in the case of
$\pi$-periodic distributions $p(\vphi)$, since in such cases centroid'
angle is not determined (the centroid is in the origin).  The subtle
however is easily overcome if one note \ci{FA01} that {\it the centroid is
a natural measure of the mean of the distribution}. This gives a hint to
define more generally the center of the packet $\vphi_c$ as solution of
the equation
\beq\lb{18}
M\vphi(\vphi_0) = \vphi_0,
\eeq
where $M\vphi(\vphi_0)$ is the limit-dependent mean of $\vphi$ given by (8a).
The examination shows that the centroid' angle $\vphi_c$, when exits, is a
solution of eq. (\ref{18}). For $\pi$-periodic distributions the centroid
is in the origin, and $\vphi_c$ remains undefined. It turned out that for
such distributions eq. (\ref{18}) has more than one solution, i.e. there
are several equivalent points $\vphi_{c,i}$. We will say that in such
cases several points $\vphi_{c,i}$ on the circle may serve as "centers of
the packet", or the packet is {\it "multi-centered"}.
If $p(\vphi+\pi/k) = p(\vphi)$, $k=1,\ldots,n$ then equation (\ref{18})
should have $2n$ different solutions $\vphi_{c,i}$, $i=1,\ldots,2n$.  For
$p_s(\vphi)$, $p_{2s}(\vphi)$ on figure 1 (and $p_{cs}(\vphi)$,
$p_{s4}(\vphi)$ on figure 2) we have solutions $\vphi_c = \pm\pi/2$,
$\vphi_c = \pm\pi/4,\pm3\pi/4$ (and $\vphi_c=0$, $\vphi_c = 0, \pi$). For
the uniform distribution eq.  (\ref{18}) degenerates to the
identity $\vphi_0=\vphi_0$, i.e. for $p_0(\vphi)$ all points on the circle
are equivalent.

The equation (\ref{18}) may be difficult for analytical handling, but
solutions can be easily found numerically, or by the following rule/anzatz:\,
$\vphi_{c,i}$ are points $\vphi_{\rm min}$ of the global minimum of
the second moment $D\vphi(\vphi_0)$, eq. (\ref{8}), as a function of
$\vphi_0$.  This means that $\Dlt^2\vphi(\vphi_{c,i})$, $i=1,\ldots,n$,
coincide, and
\beq\lb{19}
D\vphi(\vphi_{c,i}) = \,_b\Dlt^2\vphi,\quad i=1,\ldots,n.
\eeq
The rule works (is confirmed) on the example of a variety of distributions
$p(\vphi)$,  in particular on all examples in figures 1 and 2. Since the
global minimum of $D\vphi(\vphi_0)$ can be calculated invariantly in any
interval $I_{2\pi}\ni \vphi_0$ the above coincidence confirms again that
the measure $\,_c\Dlt^2\vphi$ depends solely on the state.
\vspace{3mm}

\section*{\large 4. Conclusion}

In this paper we have introduced and discussed new position uncertainty
(delocalization) measures  for a particle on the circle.  The relevant
measure properties are illustrated on several examples, where the previous
measures (based on position operators $\sin\vphi$, $\cos\vphi$, or
$\exp(i2\vphi)$) appear to be unsatisfactory.
The new measures resort on multiplication angle operator variance (see
eqs. (\ref{11}), (\ref{12}), (\ref{17})) and obey uncertainty relations of
the Schr\"odinger--Robertson type (with appropriate generalizations of the
notions of covariance and mean commutator for the angle and angular
momentum observables). The first two measures are defined as arithmetic
mean of the angle variance or as minimal value of the variance within any
$2\pi$ length interval. The latter appears to coincide with the angle
variance, calculated by integration from $\varphi_c-\pi$ to $\varphi_c +
\pi$, where $\varphi_c$ is the center of the wave packet, defined
appropriately. The values of these measures are determined solely by the
wave function $\psi(\vphi)$ of the particle.

The position and the angular momentum uncertainty measures can be used to
define delocalization measures on the phase space (here it is a cylinder
$S^1\times I\!\!R$). Such measures can be defined as a sum or as a
product of position uncertainties $\,_i\Dlt^2\vphi$, $i=a,b,c$, and
angular momentum variance $\Dlt^2l_z$. These possibilities stem
from eqs. (\ref{9}), (\ref{14}). From (\ref{9}) and
(\ref{14}) we also derive the uncertainty relations 
\beq\lb{20}
\Dlt^2l_z + \,_i\Dlt^2\vphi \,\geq\, 2|_i({\rm Im}\,G_{\vphi l_z})|,\quad
i=a,b,c.
\eeq
The counterpart of this inequality on the real line is $\Dlt^2x +
\Dlt^2p_x \geq 1$, which is minimized in the canonical CS
$|\alpha\ra$ {\it only} \ci{T00}. There are no periodic wave functions on
the circle, that precisely minimize (\ref{20}). Calculations show that
they are approximately minimized in the CS on the circle $|\xi\ra$
\ci{DG93,KRP96,GO98}: in $|\xi\ra$ the sum $\,_c\Dlt^2l_z + \,_c\Dlt^2\vphi$
attains the minimal value, which is very close to $1$. In this
sense $|\xi\ra$ are most localized states in the phase space. Let us
note, that in $|\xi\ra$ one also has $\Dlt^2(\hat{l}_z) + \Dlt^2(\hat\vphi)=1$
\ci{KR02}.

\vspace{5mm}

{\large Figure Captions} 
\vspace{3mm}

{\small {\bf Figure 1.}\,
 $\pi$-periodic, $\pi/2$-periodic and uniform distributions
on the circle $p_s(\varphi)=|\psi_s(\varphi)|^2$,
$p_{s2}(\varphi)=|\psi_{s2}(\varphi)|^2$ and $p_0(\varphi) = 
|\psi_m(\varphi)|^2$. The functional  $\tilde{\Delta}\varphi$, eq.
(5), on all these distributions takes the same maximal value of $1$,
while $\Delta^2(\hat{\varphi})$, eq. (7), takes the values $0.346$,
$\infty$ and $\infty$ respectively.
\vspace{3mm}

{\bf Figure 2.}\,
One- and two-peak $\varphi$-distributions $p_{cs}(\varphi)$,
$p_{s4}(\varphi)$, corresponding to the CS $|\xi\!=\!1\rangle$  and to state
$\psi_{s4}(\varphi) = {\mbox{\small const.}}\,(0.2+\sin^2\varphi)^2$ on the
circle.  Here $\tilde{\Delta}_{p_{cs}}\varphi \,<\,
\tilde{\Delta}_{p_{s4}}\varphi = 1$, while $\Delta^2_{p_{cs}}(\hat\varphi)
\,>\, \Delta^2_{p_{s4}}(\hat\varphi) = 0.143$. }

\newpage

\begin{figure}
\centering
\includegraphics[width=16cm,height=23cm]{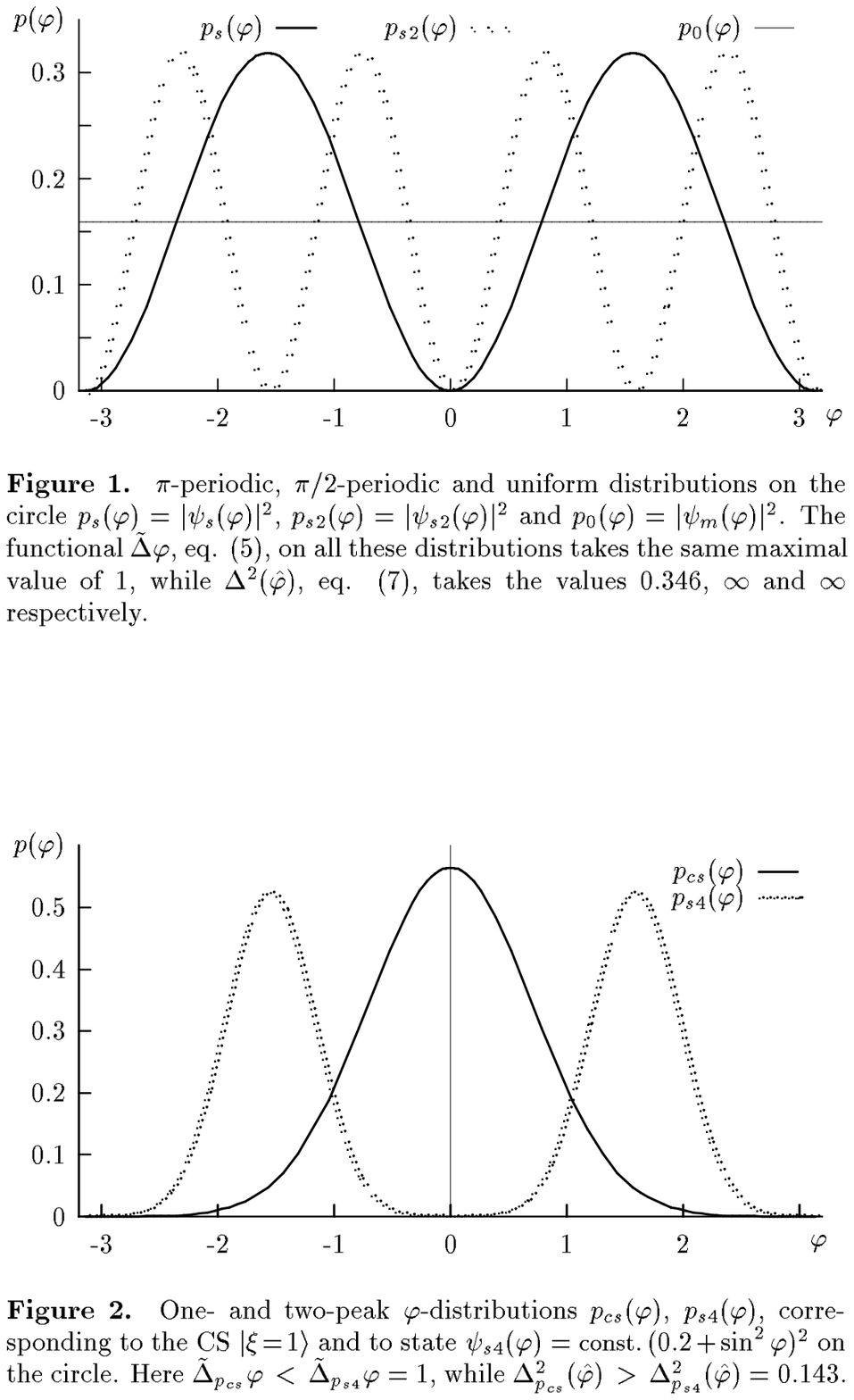}
\end{figure}

\end{document}